\documentstyle[aps,pre, preprint]{revtex}
\begin{document}

\def\wig#1{\mathrel{\hbox{\hbox to 0pt{%
    \lower.5ex\hbox{$\sim$}\hss}\raise.4ex\hbox{$#1$}}}}
\def\lsim{\wig <}
\def\gsim{\wig >}

\draft

\title {Orientational relaxation in Brownian rotors with frustrated
  interactions on a square lattice}

\author{Sung Jong Lee$^1$ and Bongsoo Kim$^2$}

\address{ $^{1}$ Department of Physics,   
  The University of Suwon, Hwasung-Gun, Kyunggi-Do 445-890, Korea }

\medskip

\address{ $^{2}$ Department of Physics,   
  Changwon National University, Changwon 641-773, Korea }

\medskip

\maketitle
\begin{abstract}

We present simulation results on the equilibrium relaxation 
of Brownian planar rotors based on a uniformly frustrated XY model
on a square lattice.  
The rotational relaxation exhibits typical dynamic features of fragile
supercooled liquids including the two-step relaxation. We observe a 
dynamic cross-over from high temperature regime with Arrhenius behavior
to low temperaure regime with temerature-dependent activation energy.
A consistent picture for the observed slow dynamics can be given in terms of
caging effect and thermal activation across potential
barriers in the energy landscapes.

\end{abstract}


\section{Introduction}

Last decade or so have witnessed significant advances in our 
understanding of the underlying mechanism for the slow dynamics of supercooled 
liquids approaching the glass transition \cite{sciean} .  
The development of mode-coupling theory of supercooled liquids \cite{mct} 
and extensive  experiments and computer simulations \cite{ykis} have played crucial 
roles in such advances.
Some efforts have also been devoted to devise model systems (even though
somewhat artificial)  \cite{model} which show glassy behavior 
similar to that of supercooled liquids. 
One line of research along this direction is to find (lattice) model
systems with no quenched disorder but some intrinsic frustration built into the 
model, which may exhibit glassy relaxations \cite{bm,fh,cfi,kl97}.  

One can imagine that there may exist a common microscopic mechanism which 
underlies the observed similarities in the relaxations of model
systems and real supercooled liquids. 
This possibility is made more plausible  by the universal scaling property 
observed in the dielectric susceptibilities of a variety of supercooled liquids 
\cite{nagel} and some plastic (glassy) crystal \cite{suga,birge,bll}.  
Here in this work, we address the question of this  possible common mechanism 
by investigating the equilibrium orientational relaxation of planar Brownian rotors 
whose interaction is prescribed by that of uniformly frustrated XY (UFXY) models
with dense frustration,  which is a prime example of non-randomly frustrated systems 
\cite{nelson} characterized by complex degeneracy of ground states and many metastable states.

While a recent simulation \cite{kl97} of the present authors
deals with the relaxation of the vortex charge density for a purely dissipative
dynamics, here we examine directly the orientational relaxation with finite rotational
inertia, which offers more transparent views on the origin of the observed 
slow relaxation.
Also, due to the one-dimensional nature of the phase of the planar
 rotors, it is convenient to probe the properties of the angular motions of the rotors of
the system.
 We find that, by including  phenomenological
rotational inertia in the dynamic equation for the rotors, the orientational
correlation exhibits a two-step relaxation, which is analogous to the 
(fast) $\beta$ and $\alpha$ relaxations of supercooled liquids.
Mean square angular displacement (MSAD) exhibits three stage behavior, 
i.e., the early time ballistic, intermediate sub-diffusive, and late time diffusive
regimes, which is argued to be consistent with the picture of
the cage effect and long-time activated dynamics for the motion of the rotors. 
It is shown that there exist two dynamically distinct regimes: a high temperature
regime where the dynamics is governed by a temperature-independent activation 
energy, and
a low temperature regime, in which the activation energy increases with decreasing 
temperature, which is interpreted as arising from complex energy landscapes 
\cite{mgold,still} probed by the system in the low temperature regime. 

\section{Dynamic model and Simulation method}

We consider the following Langevin dynamics for a collection of planar rotors
on a square lattice
\begin{equation}
I \dot{\omega}_i(t)+\gamma \omega_i(t) = - \frac{\partial V(\{\theta\})}{\partial \theta_i(t)}
+\eta_i(t)
\label{eqn:n1}
\end{equation}
where $I$ is the moment of inertia, $\omega_i(t)\equiv \dot{\theta}_i(t)$ the
angular velocity of the rotor at site $i$, $\gamma$ the damping constant, and 
$\eta_i(t)$ the thermal noise.
The equation (\ref{eqn:n1}) describes the Brownian motion of rotors
subject to the interaction potential energy $V(\{\theta\})$.
The thermal noise $\eta_i(t)$ is given by a gaussian random variable 
\begin{eqnarray}
<\eta_i(t)>&=&0 \nonumber \\
< \eta_i(t) \eta_j(t^{\prime})>&=&2\gamma T\delta_{ij} \delta(t-t^{\prime}) 
\label{eqn:n2}
\end{eqnarray}
where the Boltzmann constant $k_B$ is set equal to unity.
The variance of the noise in (\ref{eqn:n2}) ensures that  
the system at temperature $T$ evolves toward the equilibrium state 
whose properties are governed by the Boltzmann distribution
$\exp(-E(\{\theta\},\{\omega\})/T)$ where the energy $E(\{\theta\},\{\omega\})$
is given by $E(\{\theta\},\{\omega\})=I\sum_i \omega_i^2/2+V(\{\theta\})$.

Here we chose the potential energy $V(\{\theta\})$ as the energy of 
the two dimensional UFXY model on a square lattice which takes the form \cite{tj}  
\begin{equation}
V(\{\theta\})=-J \sum_{(ij)} \cos(\theta_i-\theta_j-A_{ij})
\label{eqn:n3}
\end{equation}
where $J$ is the coupling constant and $(ij)$ denotes nearest neighbor pairs. 
The bond angles $A_{ij}$ satisfy the constraint
\begin{equation}
\sum_{i,j \in P} A_{ij}=2 \pi f
\label{eqn:n4}
\end{equation}
where the sum is over $(i,j)$ belonging to the unit plaquette $P$
causing competing interactions (frustration) between the rotors.
Here, $f$ is called the frustration parameter of the system.
 
A convenient choice for $A_{ij}$ is the Landau gauge which is given by $A_{ij}=0$ 
for every horizontal bond and $A_{ij}=\pm 2\pi f x_i $ 
for the vertical bond directed upward (downward) with $x_i$ being 
the $x$-coordinate of the site $i$.
It can be readily checked that this choice of the bond angles obeys the condition
(\ref{eqn:n4}).
Due to the invarince of the Hamiltonian (\ref{eqn:n1}) under $f \rightarrow f+1$ 
and $f \rightarrow -f$, we need to consider the values of $f$ only over the
range $[0, 1/2]$.  
A physical realization of this model can be found in the two dimensional
square array of Josephson junctions under a uniform perpendicular 
magnetic field. In this situation, the bond angle $A_{ij}$ is  identified with the line 
integral of the vector potential ${\bf A}$ of the transverse magnetic field:
$A_{ij} = (2\pi/\Phi_0)\int_i^j {\bf A} \cdot d {\bf l}$
where $\Phi_0$ is the flux quantum $\Phi_0 \equiv h c/2e $ 
per unit plaquette. With this identification the strength of magnetic field $B$ is 
given by $Ba^2=f \Phi_0$ where $a$ is the lattice constant.

The UFXY model can be mapped \cite{vill} onto that of a lattice Coulomb gas
with charges of magnitude $(n-f)$, $n=0, \pm 1, \pm 2, \cdots $, 
where charges correspond to phase-vortices with suitably defined vorticity around 
the plaquettes. 
The lowest excitation consists of  charges with magnitudes $1-f$ and $-f$, respectively. 
The charge neutrality condition then implies that 
the number density of positive charges is equal to $f$.
For the case of $f=0$, the well-known Kosterlitz-Thouless transition \cite{kt}
occurs via vortex-antivortex unbinding at a finite temperature. 
Except for this case of unfrustrated XY model, 
the equilibrium nature  and associated phase transitions of  these systems are 
not very well understood even for the next simplest case of $f=1/2$, 
the so-called full frustrated XY model \cite{gkn}.
For example, the ground state configurations for the case of general $f=p/q$ 
($p$ and $q$ are relative primes) are not known \cite{hals1,myc} 
except for some low order rational values of $f$, 
such as $f=1/2$, $1/3$, $2/5$, $3/8$, etc, where staircase type of 
ground state configurations are known analytically \cite{hals2,myc}. 

As $q$ becomes large (the limit of irrational frustration), due to the
complexity of the degeneracy of the system and long equilibration time,
 it is quite a difficult task to 
analyze the nature of the low temperature phase of the system. And, inspite of
recent claim by Denniston and Tang \cite{dt} that there exist a first order transition 
 near $T_c \simeq 0.13J$, in  the case of $f=1-g$, ($g$ being the golden-mean 
ratio $g=(\sqrt{5}-1)/2 \simeq 0.618$), it is fair to say that the low temperature 
phase is not completely understood yet. On the other hand, since it is clear 
that many metastable states are possible due to the dense frustration,
one can expect that Brownian dynamics (\ref{eqn:n1}) with the potential 
energy (\ref{eqn:n3}) may generate a slow relaxation where trapping of the configurations
in deep metastable minima and thermal activation across the potential 
barriers play a crucial role.
Note that there is no intrinsic disorder in the present system, 
which distinguishes itself from a spin glass system where 
both intrinsic disorder and frustration are considered to be essential \cite{biy}.

With the potential energy (\ref{eqn:n3}), the Langevin equation is explicitly given by
\begin{equation}
I \dot{\omega}_i(t)+\gamma \omega_i(t) = -J \sum_j \sin(\theta_i-\theta_j-A_{ij})
+\eta_i(t)
\label{eqn:n5}
\end{equation}
We integrate the equation (\ref{eqn:n5}) in time, starting from
random initial conditions $\{\theta_i(0)\}$ and $\{\omega_i(0)\}$
using an Euler algorithm  on a square lattice of linear size $N=34$, 
In our simulations, we used $I=1.5$, $\gamma =1$, $J=1$ and $f=13/34$, 
which is a Fibonacci approximant to $f=1-g$. 
 Periodic boundary conditions are employed for both spatial directions. 
The results were averaged over $150 \sim 1000$ different random initial configurations, 
depending on the quenching temperature.
As for the integration time step, we used $dt = 0.05$ in the 
dimensionless unit of time.  No essential difference could be found in the 
results when compared with those obtained by using $dt =0.01$.  

\section{Results and Discussions}

In order to probe the orientational relaxation of the system we first computed
the on-site auto-correlation function for the planar spins
\begin{equation}
C_R(t)= \frac{1}{N^2} \left< \sum_{i=1}^{N^2}\cos(\theta_i(0)-\theta_i(t)) \right>
\label{eqn:n6}
\end{equation}
where the bracket $<\cdots>$ in (\ref{eqn:n6}) represents an average over different random 
initial configurations. 
In this work we focus only on the lowest order correlation even though one may also measure 
the higher order correlations, as was done in recent molecular dynamics simulations
 \cite{wl,sgtc96,kks97}.

Shown in Fig.~1 is the on-site auto-correlation function $C_{R}(t)$. 
The relaxation continuously slows down as the temperature is lowered. 
In order to characterize the slowing down of the relaxation, 
one can define a characteristic relaxation time $\tau_R(T)$ as $C_R(\tau_R)=1/e$. 
The temperature dependence of $\tau_R(T)$ is shown in the inset of Fig.~1. 
It exhibits an Arrhenius behavior
at high temperatures, while at low temperatures ($T < 0.20$) it
shows a non-Arrhenius behavior, which can be well fitted by the Vogel-Tamman-Fulcher form
$\tau_R(T) = \tau_0 \exp[DT_0/(T-T_0)]$ with $\tau_0 \simeq 9.92$, $T_0 \simeq 0.08$, and
$D \simeq 3.58$ \cite{torell}.  Similar non-Arrhenius behavior was observed 
in the vorticity relaxation as well \cite{kl97}.

An interesting feature of the rotational relaxation is 
that it exhibits a two-step relaxation, a very fast relaxation 
(up to $t\simeq 3$ for $T=0.13 J$, the lowest temperature probed)
and a slow relaxation following the fast relaxation.
The earliest part of the fast relaxation is expected to be well described by
the free rotation of the rotors $I\dot{\omega}_i(t)+\gamma \omega_i(t)=0$.
For the time range where $t \ll I$, the inertial term is dominant and 
hence $\theta_i(t)-\theta_i(0)\simeq \omega_i(0)t$.
It is then easy to show that the relaxation is given by $C_R(t) \simeq 1-(T/2I)t^2$ 
using the equipartition theorem $\bigl<\omega^2\bigr>=T/I$.

The long-time part of the slow relaxation can be well fitted by the stretched
exponential form $C_R(t)=C_0 \exp[-C_1 (t/\tau_R)^{\beta}]$
($C_1 = 1 + \ln C_0$ due to the definition of $\tau_R$), shown in Fig.~2.
We find that the exponent $\beta$ varies with temperature: it 
decreases as the temperature is lowered, as shown below in the inset of Fig.~3.
It is interesting to note that at low temperatures ($T \leq 0.2$) the
short time part of the slow relaxation shows a deviation from its stretched
exponential fit and the time region for this deviation tends to extend over
longer time regions with lowering temperature. We have fitted this region
with a power law decay known as the von-Schweider relaxation \cite{gs} 
$C_R(t)=C_2 - C_3 t^b$ where the exponent $b$ also varies with
temperature (see the inset of Fig.~3).
We now examine the scaling behavior of the rotational relaxation.
Shown in Fig.~3 is $C_R(t)$ versus the rescaled time $t/\tau_R(T)$. 
Obviously the earliest part of the relaxation does not obey the scaling 
since faster time scale (the inverse of the inertia which is temperature 
independent) is involved in this regime. We also observe that the
time-temperature superposition of the relaxation function is systematically 
violated in the late (slow) part of the relaxation, especially at low 
temperatures. 
This breakdown of the scaling is consistent with the fact that the
two exponents $b$ and $\beta$ vary with temperature.

It would be interesting to examine the response function corresponding to 
the orientational correlation function $C_R(t)$. 
The response function in the frequency ($\nu$) domain can be  defined as 
(via fluctutation dissipation theorem) 
$\chi^{''}(\nu)=2 \pi \nu \int_0^{\infty} dt \cos(2\pi \nu t)C_R(t)$.
Fig.~4 shows $\chi^{''}(\nu)$ versus $\nu$ in a semi-log plot.
We see that there exist two peaks, the low-frequency $\alpha$ peak
and the high-frequency peak (microscopic peak). 
As the temperature is lowered, the $\alpha$-peak moves to lower frequency, indicating
the slowing-down of the reorientational relaxation. At the same time, the maximum value
of $\chi^{''}(\nu)$, which is analogous to the Debye-Waller factor,
continuously decreases, and the $\alpha$-spectrum becomes broadened 
as the temperature is lowered. We also note that as the temperature is lowered
a minimum of the spectrum is slowly developed. 
All these features in the frequency spectrum of the orientational relaxation 
is qualitatively quite similar to the recent broad-band dielectric 
susceptibility measurement of supercooled liquids \cite{nagel,menon,lunk}.  
According to the recent dielectric susceptibility data, the $\alpha$-spectrum  
of supercooled liquids consists of two power law regimes in the right-hand side 
of the $\alpha$-peak. 
The first power law relaxation clearly corresponds to the stretched exponential 
relaxation in time domain. In addition to this, another power law regime is observed 
in the high frequency side of the $\alpha$-spectrum. 
It is quite interesting that similar  power law 
relaxation is also observed in the high frequency part of the magnetic 
susceptibility of a spin glass system \cite{bitko}.
Although we can not better resolve the high frequency part of the $\alpha$-spectrum
of the present orientational relaxation due to the bad statistics of the
spectrum at low temperatures, we believe that our orientational relaxation spectrum
also exhibits similar two-power-law regimes in the right hand side of the $\alpha$-
peak. The reason is that, even though the long time part of $C_R(t)$ can be well 
fitted by a stretched exponential function, the regime of its validity 
(for stretched exponential form) is limited to late time regime only and does not
extend to intermediate time regime where so called von-Schweidler relaxation \cite{km}
(with different exponent $b$) better fits the relaxation function. In the frequency 
domain this will correspond to two power law behavior.

In order to investigate the self-diffusion of the rotors, 
we  measured the mean squared angular displacement (MSAD) 
\begin{equation}
\bigl<(\Delta \theta (t))^2 \bigr> = \frac{1}{N}
\left<\sum_{i=1}^N (\theta_i(t)-\theta_i(0))^2 \right>
\label{eqn:n7}
\end{equation}
where the phase angle $\theta_i(t)$ is unbounded. 
Fig.~5 shows a log-log plot for the MSAD
$\bigl<(\Delta \theta (t))^2 \bigr>$ versus time $t$.
For all temperature range probed, we see that $\bigl<(\Delta \theta (t))^2\bigr> \sim
t^2$ in the early time regime, which may be called the ballistic regime.
It is expected that each rotor makes a free rotation 
in this time regime. Hence the MSAD is then given by 
 $\bigl<(\Delta \theta(t))^2\bigr> \simeq (T/I)t^2$ in the ballistic regime. 
This regime corresponds to the earliest part of the relaxation $C_R(t) \simeq 
1-(T/2I)t^2$.
For high temperatures this ballistic regime directly crosses over to 
the diffusive regime where $\bigl<(\Delta \theta (t))^2\bigr> \sim t$. 
But as the temperature is lowered,  in the intermediate time regime a sub-diffusive regime  
characterized by $\bigl<(\Delta \theta (t))^2\bigr> \sim t^{\phi}$ with $\phi < 1$ 
(for example, $\phi \simeq 0.3$ for $T=0.13 J$)
starts to appear and extends over more than two decades of time at 
the lowest temperature probed ($T=0.13J$).
The sub-diffusive regime sets in at the same time $t \approx 2$ for all
temperatures. In this regime the rotational motion is significantly hindered. 
This can be directly seen in Fig.~6 which shows the angular displacements 
$\Delta \theta_i(t) \equiv \theta_i(t)-\theta_i(0)$ at 
some representative sites at $T=0.15J$.  
We clearly see from this figure that for all these phase angles  
the rotational motion looks almost frozen for more than a few thousand time 
units.
This strongly indicates that the system is stuck in a particular configuration 
among many possible metastable states.
The rotor then executes a local vibrational motion only, which
corresponds to the caging in the dynamics of real supercooled liquids.
%In this regime, we may approximate the potential energy (\ref{eqn:n3})
%by a harmonic expansion around the (above-mentioned) metastable configuration.
%we can no longer use the harmonic approximation
At longer time scales, however, the local rotors can execute full rotations 
via activated tunneling through
the potential barriers, showing occasional abrupt rotational motions, 
as shown in Fig.~6.  Similar jump motions have been observed in MD simulations of 
soft-sphere mixtures \cite{hiwa}, binary Lennard-Jones \cite{gw}, and the
colloidal glass \cite{sood}.
Also, neighboring rotors can execute collective rotations,  
 thereby slowly rearranging the whole phase configurations. 
This stage will correspond to the slow part of $C_R(t)$.
%In order to make quantitative comparison, one could perform such computations of
%harmonic approximation,  as was recently done in the MD simulations of supercooled
%water \cite{st97}.
This entire time evolution of the self rotational motion 
is qualtitatively the same as that observed in MD simulations of the orientational 
relaxation of molecular supercooled liquids \cite{kks97}.

The rotational diffusion constant $D_R(T)$ can be obtained by 
the slope of the MSAD versus $t$ in the long time limit where
MSAD exhibits diffusive behavior $\bigl< (\Delta \theta(t))^2 \bigr> =2D_R(T)t$. 
As shown in Fig.~7, at high temperatures the rotational diffusion 
constant  exhibits an Arrhenius behavior, which is well fitted by
$D_R(T)=D_0 \exp(-\Delta E/T)$ with $D_0 \simeq 0.68 $ and the temperature
independent activation energy $\Delta E \simeq 0.87 J$.
As the temperature is lowered, however, $D_R(T)$ shows a strong deviation 
from the Arrhenius behavior. This behavior implies  that 
the long time dynamics in the high temperature regime is governed by
activation barriers whose average height does not depend on temperature.
In the low temperature regime, the rotors explore deeper valleys in the 
potential energy landscapes whose depth increases as the temperature
decreases, giving rise to the non-Arrhenius behavior of the relaxation
time \cite{sds}. 

It was observed in some experiments of supercooled liquids \cite{fgsf} that while 
both translational and rotational diffusion constants are proportional to the 
inverse of viscosity at high temperatures, the decrease of the translational 
diffusion constant is less dramatic than the inverse of viscosity at low temperatures.
The rotational diffusion constant, on the other hand, is still proportional to 
the inverse of viscosity at low temperatures down to the glass transition.
This relative enhancement of the translational self-diffusion
is  also revealed in recent simulations of supercooled liquids \cite{peharro,onuki} 
and the lattice model systems \cite{nico,lklcg}.
Here we compared the temperature dependences of the two time scales $1/D_R(T)$ and 
$\tau_R(T)$. Shown in the inset of Fig.~7 is a plot for $D_R(T)\tau_R(T)$ versus
$T$. 
Since the product $D_R(T)\tau_R(T)$ in the plot is measured to be nearly contant
down to $T=0.20 J$, the two time scales are observed to be proportional to each 
other, i.e., $\tau_R(T) \sim D_R(T)^{-1}$ up to $T=0.20 J$. 
The data points below $0.20 J$ tend to deviate from this proportionality, 
indicating more rapid decrease (rather than enhancement) of the rotational 
diffusion constant. However, it is not clear to us whether this anomalous behavior
 is a genuine feature of the present model or not.

We have also measured the normalized angular velocity auto-correlation function (AVCF)
\begin{equation}
C_{AV}(t)=\frac{\bigl< \sum_{i=1}^{N^2}\omega_i(0)\omega_i(t)\bigr>}
{\bigl< \sum_{i=1}^{N^2}\omega_i^2(0)\bigr>}.
\label{eqn:n8}
\end{equation}
In the absence of the interaction between rotors, $C_{AV}(t)$ can be easily obtained as
$C_{AV}(t)=\exp(-\gamma t/I)$. With interaction, as shown in Fig.8, the AVCF shows 
a strongly damped oscillatory motion. As the temperature is lowered, the amplitude of 
oscillation becomes enhanced. This behavior strongly indicates that the rotors
execute angular rattlings in `cages' \cite{mount}.  

For purely gaussian distribution of the angular displacements,
it is easy to show that the rotational correlation
function $C_R(t)$ can be expressed in terms of the mean square angular displacement
$\bigl<(\Delta \theta(t))^2\bigr>$ as $C^{(G)}_R(t) \equiv 
\exp(-\bigl<(\Delta \theta(t))^2\bigr>/2)$. 
Shown in Fig.~9 is the comparison of 
the rotational correlation function $C_R(t)$ and its gaussian approximation $C^{(G)}_R(t)$. 
We find that $C_R(t)$ exhibits a good agreement with the gaussian approximation
in the early time regime whereas it shows a considerable deviation from the
gaussian approximation in the late time regime.
In order to characterize the non-gaussian nature of the distribution of displacements, 
the non-gaussian parameter has often been used in simulations of 
supercooled liquids \cite{rah,hiwa91,kob95,harro}. 
Here we measure the same quantity for the angular displacements, which is defined as
\begin{equation}
\alpha_2(t) = \frac{1}{3} 
\frac{\bigl<(\Delta \theta(t))^4\bigr>}{\bigl<(\Delta \theta(t))^2\bigr>^2}-1
\label{eqn:n9}
\end{equation}
where the factor $1/3$ comes from the one dimensional nature for the motion of the rotors.  
As shown in Fig.~10,  $\alpha_2(t)$ exhibits three time regimes of distinct
behavior, as in the MSAD.
It almost vanishes in the ballistic regime and then rapidly increases
toward its maximum in the intermediate time regime, and finally decreases again
in the long time regime. This temporal behavior is qualitatively the same as 
that  observed in some MD simulations \cite{kob95}.

As the temperature is lowered, the maximum value of $\alpha_2(t)$ rapidly increases,
and at the same time, the time regime where $\alpha_2(t)$ increases are extended, 
indicating  strong non-gaussian nature of the rotational motion in this regime.
This regime corresponds to the sub-diffusive regime in the time dependence of the 
MSAD shown in Fig.~5.  
It is expected that $\alpha_2(t)$ eventually decays to
zero since, for pure diffusion, the gaussian distribution is expected 
for the angular displacement.

\section{Summary}

We have shown that the relaxation of a phenomenological
Brownian rotors based on densely frustrated XY model Hamiltonian exhibits a slow
dynamics which is remarkably similar to
the relaxation of fragile supercooled liquids.
We find that there exist a dynamic cross-over from high temperature regime where
the dynamics can be described by temperature-independent activation energy, 
and low temperature regime where non-Arrhenius behavior sets in, which can
be attributed to the dynamic characteristics of the system probing deeper 
valleys in the potential energy landscapes with increasing height of the 
activation energy barrier.
The caging in the metastable minima and thermal activation
across potential barriers in the energy landscapes may provide
the underlying physical origin for the similarity in the slow dynamic 
behavior of the present model system and that of real fragile supercooled liquids.
It would be very interesting to quantitatively characterize the
metastable states present in the system such as finding the local minima 
and densities of metastable states. In this regard, it would also be very 
instructive to examine how the dynamic features change as the value of the
frustration parameter $f$ is varied. 
 We can also consider Newtonian dynamics version of
 our system  and compare with Langevin dynamics \cite{roux,gkb}, which may
 provide further insight into these questions.  
We will undertake further study along these directions in the near future.

We thank  Kyozi Kawasaki, Sidney  Nagel, and Peter Lunkenheimer for valuable discussions. 
This work was supported 
by BSRI (BSRI 98-2412) and by SERI, Korea through CRAY R{\&}D 98. 
We also acknowledge the generous allocation of computing time 
from the Supercomputing Center at Tong-Myung Institute of Technology.

\newpage

\centerline {\bf FIGURE CAPTIONS}

\renewcommand{\theenumi}{Fig.~1}
\begin{enumerate}
\item
    The rotational auto-correlation functions $C_R(t)$ versus time $t$ (in 
    dimensionless units with $\gamma = 1$ and $ J=1$ ) 
    for temperatures $T/J = 0.5$, $0.4$, $0.3$, $0.25$, $0.2$, $0.17$, $0.15$,
      $0.14$,  $0.13$.
    Inset: An Arrhenius plot for the characteristic relaxation time defined as
     $C(\tau_R(T)) \equiv 1/e$, where solid line is a Vogel-Tamman-Fulcher fit at low 
     temperature regime (see the text).
\end{enumerate}

\renewcommand{\theenumi}{Fig.~2}
\begin{enumerate}
\item
   Stretched exponential fit (dashed lines) to the long time part of the
   autocorrelation functions (for the same temperatures as in Fig.~1). Time $t$
   is measured in the same dimensionless units as in Fig.~1. 
\end{enumerate}

\renewcommand{\theenumi}{Fig.~3}
\begin{enumerate}
\item
    Rotational autocorrelation functions $C_R(t)$ versus the  rescaled time 
    $t/ \tau_R (T)$. Note that the time-temperature superposition is systematically
    violated.
    The inset shows the temperature dependence of the exponents $b(T)$ and $\beta(T)$ 
    characterizing the slow part of the correlation function $C_R(t)$.
\end{enumerate}

\renewcommand{\theenumi}{Fig.~4}
\begin{enumerate}
\item
    Dynamic response function $\chi^{''}(\nu)$ corresponding to the rotational 
    relaxation versus frequency $\nu$ for temperatures
    $T = 0.5$, $0.4$, $0.3$, $0.25$, $0.2$, $0.17$, $0.15$. In addition to
    the microscopic peak, one can clearly see the development of 
    $\beta$-minimum (as the temperature is lowered), decrease of the height of the 
    $\alpha$ peak and broadening of the width of the $\alpha$ peak.  
\end{enumerate}

\renewcommand{\theenumi}{Fig.~5}
\begin{enumerate}
\item
    Mean squared angular displacement $\bigl<(\Delta \theta (t))^2 \bigr>$
    versus time $t$ (in dimensionless units) for the same temperatures as 
    in Fig.~1. At the lowest temperature probed ($T=0.13 J$), sub-diffusive 
    regime extends over more than two decades.
\end{enumerate}

\renewcommand{\theenumi}{Fig.~6}
\begin{enumerate}
\item
    Angular displacement $\Delta \theta_i(t)$ versus time $t$ 
    (in dimensionless units) at some chosen lattice sites for $T=0.15 J$. 
    Rotational caging effect and occasional jump motions are exhibited.
\end{enumerate}

\renewcommand{\theenumi}{Fig.~7}
\begin{enumerate}
                                                                 
\item
    An Arrhenius plot for the rotational diffusion constant $D_R(T)$.
    We can see a crossover from high temperature regime with Arrhenius 
    behavior to low temperature regime with non-Arrhenius behavior.
    The inset shows an anomalous deviation from the Stokes-Einstein relation
    by plotting the product $D_R (T)\tau_R (T)$ versus $T$, where we can find 
    that, at low temperaures, the coefficient of angular diffusion is smaller
    than that which would be expected from standard Stokes-Einstein relation.
\end{enumerate}
    
\renewcommand{\theenumi}{Fig.~8}
\begin{enumerate}
                                                                 
\item
   The angular velocity auto-correlation functions $C_{AV}(t)$ for $T=0.50 J$ 
   and $T=0.13 J$ ($t$ in dimensionless units). 
   For comparison, dotted line represents exponential relaxation
   corresponding to the situation where the potentials are neglected. 
   One can see a strong rotational cage effect indicated by the oscillating tail of 
   $C_{AV}(t)$.
\end{enumerate}
    
\renewcommand{\theenumi}{Fig.~9}
\begin{enumerate}
\item
    The rotational autocorrelation functions versus time $t$ (in dimensionless 
    units) for temperatures $T/J = 0.5$, $0.3$,  $0.17$, $0.14$, and $0.13$ 
    together with Gaussian approximation results (dotted lines). 
    Systematic deviations are seen at late time stage.
\end{enumerate}

\renewcommand{\theenumi}{Fig.~10}
\begin{enumerate}
\item
    Nongaussian parameter versus time $t$ (in dimensionless units) for the 
    same temperatures as in Fig.~1.
\end{enumerate}

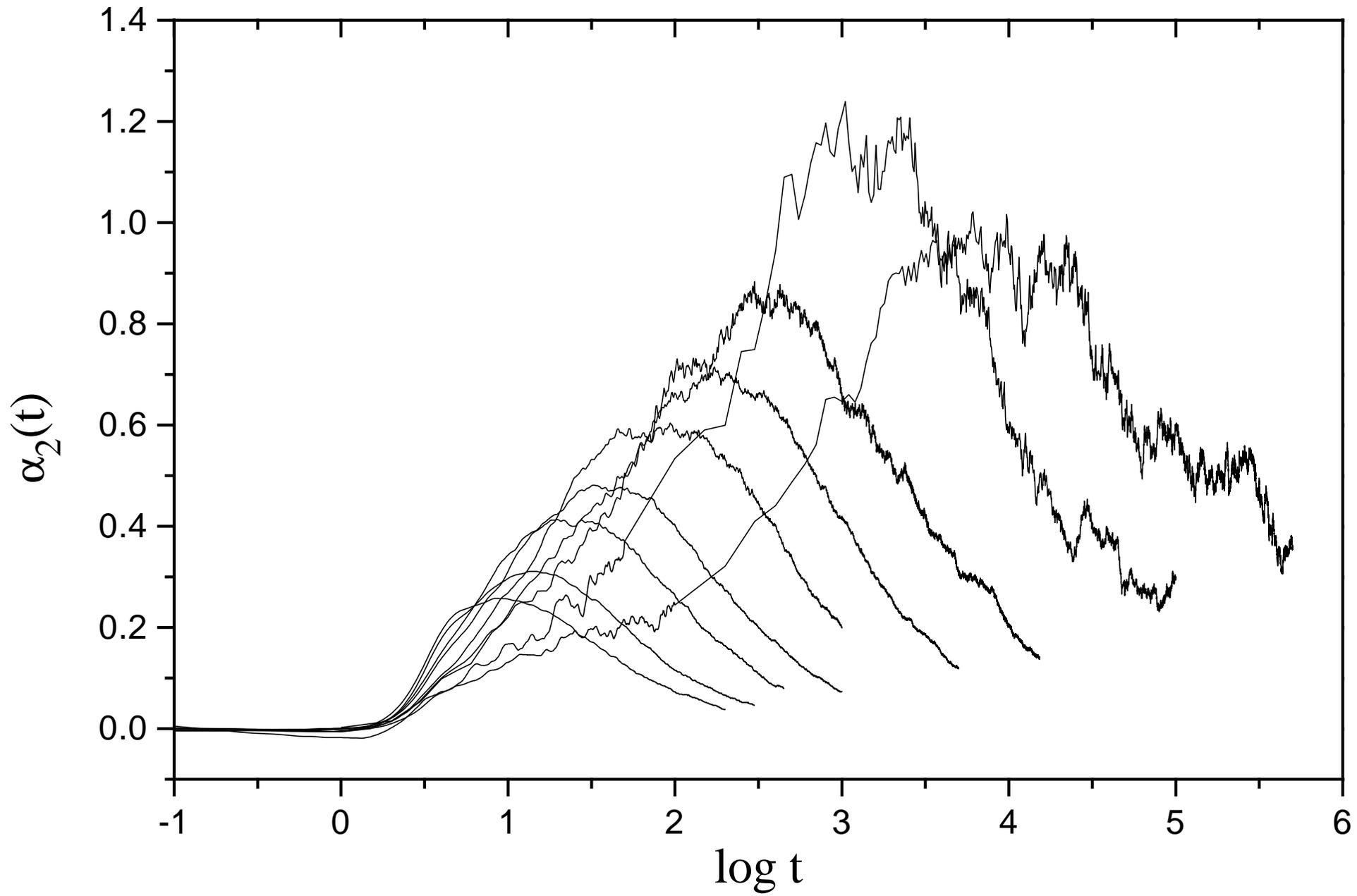
\end{document}